\documentclass{aa}  
\usepackage[dvipsnames]{xcolor}

\usepackage{hyperref}
\usepackage[normalem]{ulem}
\usepackage{float}
\usepackage[bottom]{footmisc}
\usepackage{twoopt}
\usepackage{natbib}
\usepackage{graphicx}
\usepackage{placeins}
\usepackage{mathtools}
\usepackage{amsmath}
\usepackage{bigints}
\hypersetup{colorlinks=true,
linkcolor=blue,
citecolor=blue,
bookmarksnumbered=true,%
bookmarksopen=true
}
\usepackage{enumitem,booktabs,cfr-lm}
\usepackage[referable]{threeparttablex}
\renewlist{tablenotes}{enumerate}{1}
\makeatletter
\setlist[tablenotes]{label=\tnote{\alph*},ref=\alph*,itemsep=\z@,topsep=\z@skip,partopsep=\z@skip,parsep=\z@,itemindent=\z@,labelindent=\tabcolsep,labelsep=.2em,leftmargin=*,align=left,before={\footnotesize}}
\makeatother

\usepackage{txfonts}

\begin{document}

   \title{The dynamical lineage of isolated, HI-rich ultra-diffuse galaxies}

   \author{Nilanjana Nandi\inst{1}\fnmsep\thanks{nandi.nilanjana154@gmail.com},
          Arunima Banerjee\inst{1}\fnmsep\thanks{arunima@iisertirupati.ac.in}
          \and
          Ganesh Narayanan\inst{1}
          }

   \institute{
   \inst{1}Indian Institute of Science Education and Research, Tirupati 517507, India
   }

   \date{Received May 21, 2024; Accepted November 23, 2024}

 
  \abstract
 {Ultra-diffuse galaxies (UDGs) exhibit morphological similarities with other low-luminosity galaxies indicating a possible evolutionary connection.}
{We investigate for common dynamical characteristics of isolated, HI-rich UDGs with other low luminosity field galaxies, namely the low surface brightness galaxies (LSBs) and the dwarf irregulars (dIrrs).}
{We consider samples of each of UDGs, LSBs and dIrrs. We first obtain scaling relations involving mass and structural parameters for the LSB and dIrr samples respectively, and superpose the UDGs on them. We then carry out a two-sample Anderson-Darling test to analyse whether the UDGs belong to the population of the LSBs or the dIrrs. 
We next construct distribution function-based stellar-dynamical models of these galaxies to determine their kinematical parameters. We follow up with the Mann-Whitney U-test to determine if our LSB, dIrr and UDG samples belong to different parent populations so far as kinematics is concerned. Finally, we do principal component analyses involving both structural and kinematical parameters to identify the key properties accounting for the variance in the data for the respective galaxy populations.}
{From the galaxy scaling relation studies, we note that UDGs and LSBs constitute statistically different populations. However, for UDGs and dIrrs, the null hypotheses of these statistical tests cannot be rejected for the following scaling relations: (i) stellar mass versus atomic hydrogen mass, (ii) stellar mass versus dynamical mass, and (iii) dark matter core density versus core radius mass scaling. Interestingly, the dynamical models suggest that UDGs, LSBs and dIrrs constitute different galaxy populations as reflected by their radial-to-vertical velocity dispersion, and the rotational velocity-to-total stellar velocity dispersion. Finally, we observe that the total HI and stellar mass mostly regulate the variance in the structural and kinematical data both for the UDGs and dIrrs, while the ratio of radial-to-vertical velocity dispersion, and the total HI mass dominate the same in LSBs.}
{UDGs and LSBs represent statistically different galaxy populations with respect to their mass and structural properties. However, the statistical studies do not negate the fact that the structural parameters of UDGs and dIrrs follow the same normal distributions. 
However, UDGs, dIrrs and LSBs constitute very different populations as far as their kinematical parameters are concerned. Finally, we note that the variation in the structural and kinematical data of both the UDGs and the dIrrs is mostly accounted for by their stellar mass and HI mass, whereas for the LSBs, the same is explained by the ratio of the radial-to-vertical stellar dispersion followed by the HI mass.
Thus we may conclude that the UDGs and dIrrs to share common dynamical lineage.
}

\titlerunning{Nandi, Banerjee and Narayanan}
\authorrunning{Dynamical lineage of isolated, HI-rich ultra-diffuse galaxies}
   \maketitle

\section{Introduction} \label{sec:intro}

   \noindent In the last few decades, an increasingly large number of faint galaxies have been detected, with the development of observational facilities. These low-luminosity galaxies constitute a significant fraction of the galaxy population in the nearby universe (\citealt{2015vanforty,2015vanspec}; \citealt{2016yagicat}; \citealt{2017romantrujiloUDG}; \citealt{2017shifieldUDG}; \citealt{2019prolefieldUDG}; \citealt{2023Ikedajwst}). In general, they are gas-rich and dark matter-dominated, and hence constitute ideal test-beds for galaxy formation and evolution models (\citealt{1984sandagelow,1988impeybothun,1997bothunlsb}). Ultra diffuse galaxies (UDGs) constitute a class of low-luminosity galaxies,
   and some of these galaxies are claimed to be dark matter-deficient by several studies e.g. \cite{2018natureVanDokkum}, \cite{2019vanDokkumNoDM}, \cite{2019DanieliNoDM}, \cite{2022ManceraImpact}, \cite{2024PinaNoDM} and hence enigmatic.
   The term ultra-diffuse galaxy “UDG” was coined by \cite{2015vanforty}, who defined these as galaxies with central $g-$band surface brightness $\mu_{g,0} >$ 24 mag arcsec$^{-2}$ and effective radii $R_e > 1.5$ kpc. UDGs are therefore characterised by very low central surface brightness, and relatively large stellar disk scale lengths, given their stellar masses $M_*$ in the dwarf galaxy regime ($M_*$ $\equiv$ 10$^{6-9}$ $M_\odot$). UDGs can be found in a range of environments from clusters, groups to fields, and their structure and kinematics seem to vary significantly with their environment (\citealt{2016MartinezDelgadoENV}, \citealt{2016yagicat}, \citealt{2017JanssensENV}, \citealt{2017vanderBurgENV}, \citealt{2017romantrujiloUDG}, \citealt{2018ullerENV}, \citealt{2019ForbesENV}, \citealt{2019JanssensENV}, \citealt{2019prolefieldUDG}, \citealt{2019ZaritskyENV}, \citealt{2019RomanENV}, \citealt{2020BarbosaENV}).  \\ 

   \noindent Recent observational studies have identified morphological and kinematical similarities between UDGs and other low-luminosity galaxies, thus hinting at possible evolutionary links among them. \cite{2017venholafornox} found that dwarf low surface brightness galaxies (LSBs) and UDGs in the Fornax cluster share similar structural properties, 
   possibly indicating a common genesis of these galaxy populations.  
   \cite{2019ManceraEvoUDGs} studied the structural properties of a large sample of UDGs distributed at different radii within eight nearby galaxy clusters. Curiously, they observed similarities of the distribution of the axial ratios of the UDGs with those of the late-type dwarfs and dwarf ellipticals (dEs) in the Fornax Cluster, thus suggesting a link between UDGs and dwarfs.
   While studying the kinematic properties of two field dwarf irregular galaxies (dIrrs), \cite{2017udgdirrBel} concluded that dIrrs and UDGs lie close to each other in the magnitude-effective radius space. 
   Besides, the isolated UDGs are observed to have relatively higher gas fraction compared to other gas-rich "normal" galaxies, which is similar to the trend observed in fainter and smaller dIrrs (\citealt{2003LeeDirrs}, \citealt{2017Leisalfalfa}).
   Based on their size and absolute magnitude distribution, \cite{2018Conselice} concluded that UDGs and dEs in clusters are essentially the same objects. Further, UDGs in the Coma cluster were found to lie in between dEs/dwarf lenticular galaxies (dS0s) and dwarf spheroidal galaxies (dSphs), both in the Faber-Jackson relation as well as the mass-metallicity relation \citep{2019chiliinternal}. This may suggest that UDGs constitute an intermediate evolutionary state between dE0/dS0 and dSphs. 
   These apart, by studying the stellar kinematics of KDG64, a large dSph galaxy in the M81 group, \cite{2023afanasievkdg64} concluded that it is possibly in a transitional stage between a dSph and a UDG. 
   While the formation scenario of UDGs is still debatable, modelling their structure and kinematics may enhance our understanding of their evolutionary link with other low-luminosity galaxies.
   Despite empirical evidence hinting at evolutionary connection between UDGs and other low luminosity galaxies, to our knowledge, 
   a systematic study has not been done so far to investigate the dynamical connection among them. \\
   
   \noindent Our primary aim in this paper is to investigate the dynamical lineage of isolated, HI-rich UDGs and other low luminosity samples, namely LSBs and dIrrs, and look for a possibly common origin.
   LSBs are defined as galaxies which have surface brightness lesser than the night sky, their central surface brightness in B-band ($\mu_{B,0}$) being $>$ 23 mag arcsec$^{-2}$ \citep{impey1997low}. 
   Recent studies indicate that LSBs are formed inside dark matter halos with a higher spin parameter ($\lambda >$ 0.05) (\citealt{2019MNRASLSBformation} and the references therein). Since the dark matter halo and the baryonic component have the same specific angular momentum before the formation of the galactic disk (\citealt{1983_Efstathiou_Silk,2010MoWhiteBook}), a higher specific angular momentum results in a lower stellar surface density. Therefore, LSBs can be considered to be the fall-outs of the high value of the dark matter spin parameter (\citealt{1997DalcantonLSBs}; \citealt{1998MoWhiteLSBs}; \citealt{2003BoissierLSB}; \citealt{2013KimLee}; \citealt{2019jadhavbanerjee}).
   LSBs include a broad range of galaxies based on the mass, size and morphology. However, the field LSBs, which are also the LSBs used in this study, are primarily disk-dominated late-type galaxies that can be safely classified as non-dwarfs (\citealt{de1996h}, \citealt{2000HoekLSB}, \citealt{2004O'neilLSB}).
   On the other hand, dIrrs are classified by their irregular morphology with lower intrinsic luminosities, their B-band absolute magnitude $M_B >$ -18 (\citealt{2000zeeI,2001zeeII,2020ProkhelLT}). Formation of dIrrs is attributed to two distinct formation mechanisms: 1) Collapse of primordial gas clouds 2) Tidal torquing or stirring of satellite galaxies by the gravitational potential of the host galaxy \citep{2000HunterDIRR}. 
   Field UDGs are generally considered as LSBs with lower mass but larger radial extent (for example \citealt{2019martinLSBGs,2021ProleUDGLSBs,2021BrookUDGLSBs,2022ZhouUDGLSBs}); on the other hand, they are often categorised under the same class as the dIrrs, as suggested by their morphological similarities (for example, \citealt{2017udgdirrBel,2019prolefieldUDG}). In this paper, we aim to distinguish among these different classes of galaxies in terms of their structural as well as kinematic properties, choosing suitable samples from UDG, LSB and dIrr populations.
   We first check for viable correlations between different pairs of structural parameters for each of the above galaxy populations, and in case of a correlation, note if the different populations obey the same scaling relation. We then construct the distribution function-based dynamical models of some of the galaxies in the three respective populations, employing the publicly-available stellar dynamical code Action-based Galaxy Modelling Architecture (AGAMA) \citep{vasiliev2019agama}. Finally, for each galaxy population, we do a Principal Component Analysis (PCA) (\citealt{bro2014principal}, for reference) of a set of their structural and kinematic parameters and identify the key physical properties regulating the dynamics of each population.\\

   \noindent The paper is organised as follows: In Sect. \ref{sec:scale}, we introduce galaxy scaling relations, and in Sect. \ref{sec:model}, the dynamical models. In Sect. \ref{sec:sample}, Sect. \ref{sec:input} and Sect. \ref{sec:method}, we present our galaxy samples, the source of input parameters and our methodologies respectively. We present  results and discussion in Sect. \ref{sec:result}, and finally the conclusions in Sect. \ref{conclusion}. \\
%
\section{Galaxy scaling relations}\label{sec:scale} 
   \noindent According to the current paradigm of galaxy formation, galactic disks are formed at the centres of slowly rotating dark matter halos, due to cooling and condensation of gas, followed by star formation and feedback.  Theoretical models of galaxy formation are constrained by the remarkable regularities exhibited by the observed scaling relations between different pairs of structural and kinematic properties of a galaxy population: 
   the Tully-Fisher relation, the Faber-Jackson relation, the Fall relation, and the Kennicutt-Schmidt relation are some of the well-known scaling relations observed in galaxies (see for example \citealt{1997scalingrelationbook}). However, the details of the scaling relations may tend to vary with galaxies of different morphological types, 
   possibly indicating their different evolutionary routes (\citealt{2009roychowdhuridirr}; \citealt{2009wyderGALEX}; \citealt{2019jadhavbanerjee}; \citealt{2020RomeoFromLenticular}; \citealt{2022ganeshLSBSuperthin}; \citealt{2023RomeoSpecAngMom} and the references therein). Compliance of UDGs with the scaling relations obeyed by ordinary disk galaxies has been studied earlier in the literature
   (for example \citealt{2022poulainMATLAS,2023Benavedis,2023forbq,2023HuUDGglobal} and the references therein). 
   In this context, we note that several studies have shown that the gas-rich UDGs do not follow the baryonic Tully-Fisher relation; they rotate significantly slowly compared to galaxies with similar baryonic mass. (\citealt{2019pinaoffBTFR}; \citealt{mancera2020robust}; \citealt{2020karunakaranSMUDG};  \citealt{2024DuLinAlfalfa}).
   In this work, we look for feasible scaling relations between a few other pairs of dynamical parameters for our LSB and the dIrr samples, and check if the distribution of our sample UDGs is resemblant of the LSB or the dIrr population.
\section{Dynamical model}\label{sec:model}
    \noindent As discussed, the definitions of these galaxy types are based on their optical properties like absolute magnitude and surface brightness which depend on their stellar component. For example, the dIrrs and the LSBs can be easily distinguished in terms of their optical images. On the other hand, UDGs, dIrrs and LSBs may not be very different as far as their HI parameters are concerned. In this work, we have used the observed stellar (radial stellar surface density profile) as well as HI profiles (radial HI surface density profile, HI rotation curve) to constrain the total dynamical model of a galaxy which involves a self-gravitating stellar disc subjected to the external gravitational potential of a gas disk and a dark matter halo. The underlying assumption is that the HI rotation curve is the same as the stellar one, which is not necessarily the case in general (e.g. \citealt{2012Leaman}, \citealt{2021PinaJstarMstar}, \citealt{2024Taibi}).
    \\

   \noindent We employed the publicly-available stellar dynamical code AGAMA to construct distribution function-based dynamical models for our sample galaxies. \citep{vasiliev2019agama}.  Each of our galaxy models consists of a self-gravitating stellar disk, characterized by a distribution function (DF), also subjected to the static external potentials of the gas disk and the dark matter halo. According to the strong Jeans’ Theorem, the DF $J(\mathbf{x},\mathbf{v}|\Phi)$ can be expressed as a function of the actions  that depend on the total potential $\Phi$ of the system, $\mathbf{x}$ and $\mathbf{v}$ being the position and velocity of the stellar disk particles. \\

   \noindent In this work, we have considered the stellar component of the galaxy to have a quasi-isothermal distribution function which is expressed as follows :
   \begin{equation} \label{eq:1}
       f(\mathbf{J}) = f_0({J_\phi}) \bigg(\frac{\kappa}{\sigma^2_{r,*}}\bigg) e^{-\frac{\kappa J_r}{\sigma^2_{r,*}}} \bigg(\frac{\nu}{\sigma^2_{z,*}}\bigg) e^{-\frac{\nu J_z}{\sigma^2_{z,*}}}
   \end{equation}
   Here $\kappa$ and $\nu$ represent the epicyclic frequencies in the radial and vertical directions respectively, their expressions being,
   \begin{equation} \label{eq:disk}
       \kappa \equiv \sqrt{\frac{\partial^2 \Phi}{\partial R^2} + \frac{3}{R} \frac{\partial \Phi}{\partial R}}, \;\; \nu \equiv \sqrt{\frac{\partial^2 \Phi}{\partial z^2}}
   \end{equation}
   while $\sigma_{r,*}$ and $\sigma_{z,*}$ are the stellar velocity dispersions in the corresponding directions.
   $J_r$, $J_\phi$ and $J_z$ respectively correspond to the $r$, $\phi$ and $z$ components of action of the stellar disk. 
   The actions in the radial and vertical directions are respectively $J_r = E_R/\kappa$ and $J_z = E_z/\nu$, $E_R$ representing the energy of planar motion, and $E_z$ the same for vertical motion. $J_\phi$ can be represented as $J^2_\phi = R^3\frac{\partial\Phi}{\partial R}$ and the magnitude of the total action $\mathbf{J}$ is given by $J^2 = J^2_r + J^2_\phi + J^2_z$. Here $\Phi$ is the net potential due to stars, gas and the dark matter halo, and is given by the Poisson equation as follows:

   \begin{eqnarray}\label{eq:poi}
      \nabla^2 \Phi = 4\pi G (\rho_s + \rho_g + \rho_{DM})
   \end{eqnarray}
   \noindent where $\rho_s$,  $\rho_g$ and $\rho_{DM}$ are the density of the stars, gas and  dark matter halo. In AGAMA framework, the density profile of a disk distribution is given by : 
   \begin{equation}\label{eq:diskdens}
      \rho = \Sigma_0\,\exp\bigg(-\bigg[\frac{R}{R_D}\bigg]^{\frac1n} - \frac{R_\mathrm{cut}}{R}\bigg)
      \times\left\{ \begin{array}{ll} \delta(z) & h=0 \\[1mm]
      \frac{1}{2h} \exp\big(-\big|\frac{z}{h}\big|\big) & h>0 \\[1mm]
      \frac{1}{4|h|}\, \mathrm{sech}^2\big(\big|\frac{z}{2h}\big|\big) & h<0 \end{array} \right. 
   \end{equation}
   where $\Sigma_0$ and $R_D$ respectively denote the central surface brightness and disk scale
   length. $h$ is the vertical disk scale height, and $R_{\rm{cut}}$  an inner cut-off radius that modifies the inner density distribution of the disk. S\'ersic index $n$ is taken to be, respectively, 1 and 0.5 for exponential and sech-squared (roughly) profile. We assume that the stellar and the gas components have disk-like and sech-squared radial density profiles, respectively. \\

   \noindent For modelling the dark matter halo, we use the spheroidal profile given by :
   \begin{equation}\label{eq:111}
     \rho = \rho_0  \left(\frac{r}{R_c}\right)^{-\gamma} \bigg[ 1 + \bigg(\frac{ r}{R_c}\bigg)^\alpha \bigg]^{\frac{\gamma-\beta}{\alpha}} \times \exp \bigg[-\bigg(\frac{r}{r_{\rm{cut}}}\bigg)^\xi \bigg]
   \end{equation}

   \noindent In this expression, $\rho_0$ and $R_c$ denote the dark matter core density and core radius, respectively. Besides, $\alpha$, $\beta$ and $\gamma$ are the parameters to control the steepness between the inner and outer power-law slopes, the power-law index of the outer profile and the inner profile of the dark matter distribution, respectively. By setting their values, we can recover various well-known dark matter profiles; 
   for example for $\alpha = 2$, $\beta = 2$ and $\gamma = 0$  gives the pseudo-isothermal density distribution. Finally, $r_{\rm{cut}}$ represents the length scale for truncation in the outer halo while $\xi$ characterizes the steepness of the exponential cut-off. Here we set $r_{\rm{cut}} = \infty$ and $\xi = 1/n = 1$.  \\ \\
   Finally, the theoretical profiles of the density, the mean velocity and the mean velocity dispersion of the stellar disk can be determined from the moments of the DF as given below:
   \begin{eqnarray}
       \rho_s(x) &=& \iiint d^3v \; f(J[x,v]) \label{eq:dens} \\
       \overline{v} &=& \frac{1}{\rho_s} \iiint d^3v \; v f(J[x,v]) \label{eq:vel}\\
       \overline{v^2_{ij}} &=& \frac{1}{\rho_s} \iiint d^3v \; v_i v_j \; f(J[x,v]) \\
       \sigma_{ij}^2 &=& \overline{v^2_{ij}} - \overline{v_i}\;\overline{v_j} \label{eq:disp}
    \label{eqn:svd}
   \end{eqnarray}  
   Further, the stellar surface density and the line-of-sight velocity dispersion from the model can be determined as follows:
   \begin{eqnarray}
       \Sigma(R) &=& \int_{z = -\infty}^{+\infty} dz \rho_s(R,z) \label{eq:in1}\\
       \sigma_{los} &=& \frac{1}{\Sigma(R)}\int_{z = -\infty}^{+\infty} dz \rho_s(R,z) v_z^2 \label{eq:in2}.
   \end{eqnarray}
\\

   \noindent The rotation curve (Equation \ref{eq:vel}) and stellar surface density profile (Equation \ref{eq:in1}) as determined above are matched with observations to constrain our models. \\
   
   \noindent In this problem, the DF $f(\mathbf{J})$ and the surface density profile $\Sigma(R)$ of the stellar disk are modelled, while the gas disk and the dark matter halo are taken to have a fixed profile, based on mass models available in the literature. The problem essentially reduces to determining a suitable set of parameters for the DF and the density profile consistent with Equations \ref{eq:1}, \ref{eq:poi}, \ref{eq:dens} and \ref{eq:disp}, in addition to Equations \ref{eq:in1} and \ref{eq:in2}. 
   The calculation scheme followed by AGAMA is as follows: starting with trial values of the parameters of the disk density profile (Equation \ref{eq:diskdens}), it calculates the combined potential of the stellar, gas and dark matter of the galaxy using Equation \ref{eq:poi}. Next it evaluates the three components of the action ($J_r$, $J_{\phi}$ and $J_z$), and the epicyclic frequencies ($\kappa$ and $\nu$). Using the above values, and the trial values of the stellar dispersions, it then obtains the DF using Equation \ref{eq:1}. Next it obtains the stellar density (using Equations \ref{eq:1} \& \ref{eq:dens}), and calculates the updated potential using Equation \ref{eq:poi}. The above set of steps is repeated a few times until convergence in the value of the potential is achieved. \\

   \noindent However, there is a caveat to this method. Using a quasi-isothermal DF for a realistically warm stellar disk may lead to density and velocity dispersion values not quite in agreement with the observational constraints at radii smaller than the disc scale length. However, several modifications are introduced in AGAMA to address this issue : (1) the use of a linear combination of actions, (2) assumption of a different functional form of velocity dispersion, (3) modulation of the quasi-isothermal DF with a multiplicative factor. \textcolor{black}{The potential is recomputed iteratively such that it is consistent with the density distribution, which circumvents the problem and models these profiles as closely as possible (see section 4.3 of \citealt{vasiliev2019agama}). In this study, we allow a maximum mismatch of 30\% between the model-obtained and observed stellar central surface densities.} \\

   \noindent \textcolor{black}{In principle, one could improve the model further by exploring alternative profiles for the stellar disc or its DF. In fact, AGAMA offers the choice of another DF for discs, namely the exponential distribution function. However, it is applicable only to galaxies with flat rotation curves. Since all our sample galaxies have slowly-rising rotation curves, we cannot opt for an exponential DF. Therefore, we stick to a quasi-isothermal DF for obtaining the galaxy models.}
\begin{figure*}         
   \hspace*{-0.8cm}
\minipage{0.38\textwidth}
  \includegraphics[width=0.9\linewidth]{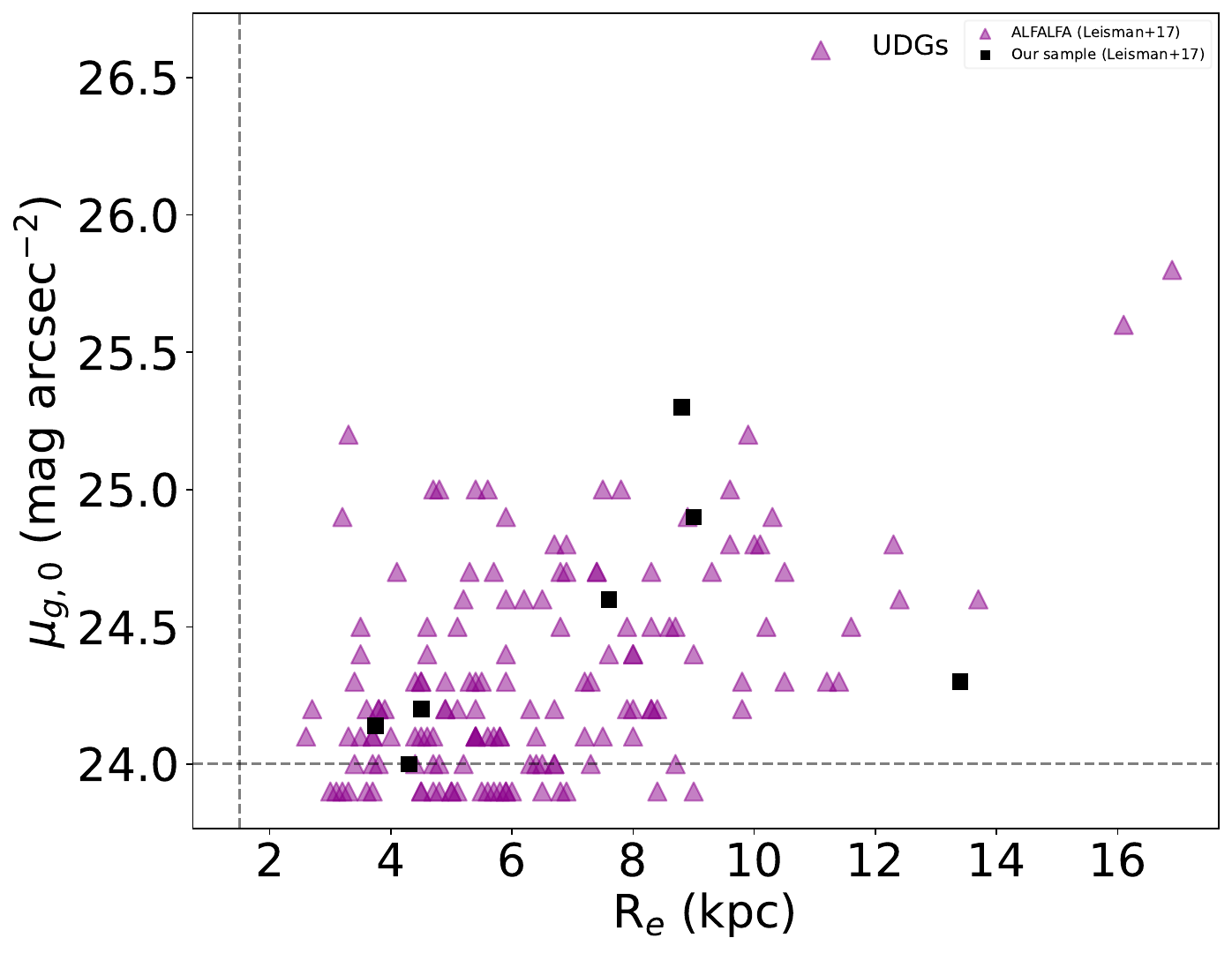}
\endminipage
\hspace*{-.5cm}
\minipage{0.38\textwidth}
  \includegraphics[width=0.9\linewidth]{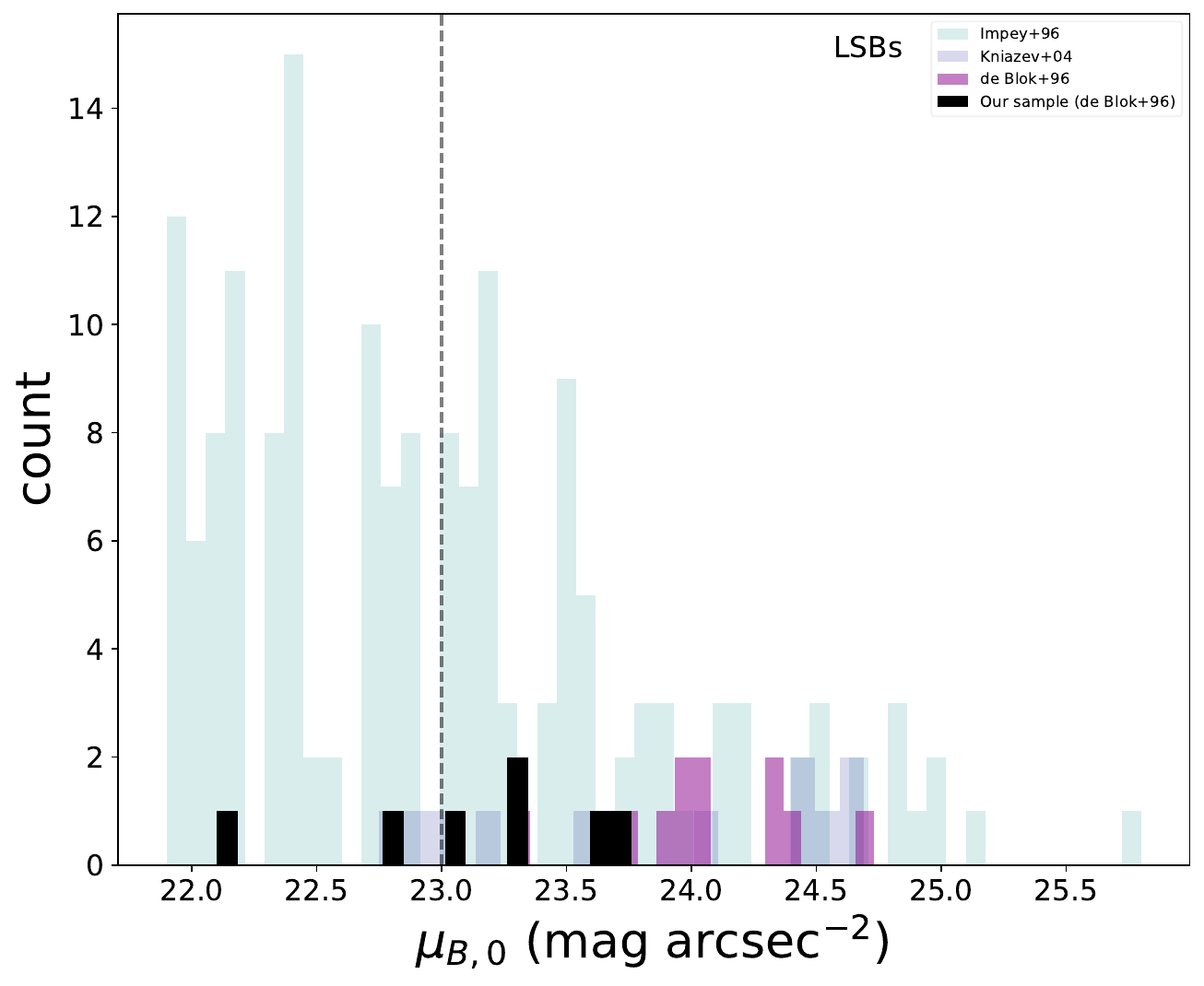}
\endminipage
\hspace*{-.5cm}
\minipage{0.38\textwidth}%
  \includegraphics[width=0.9\linewidth]{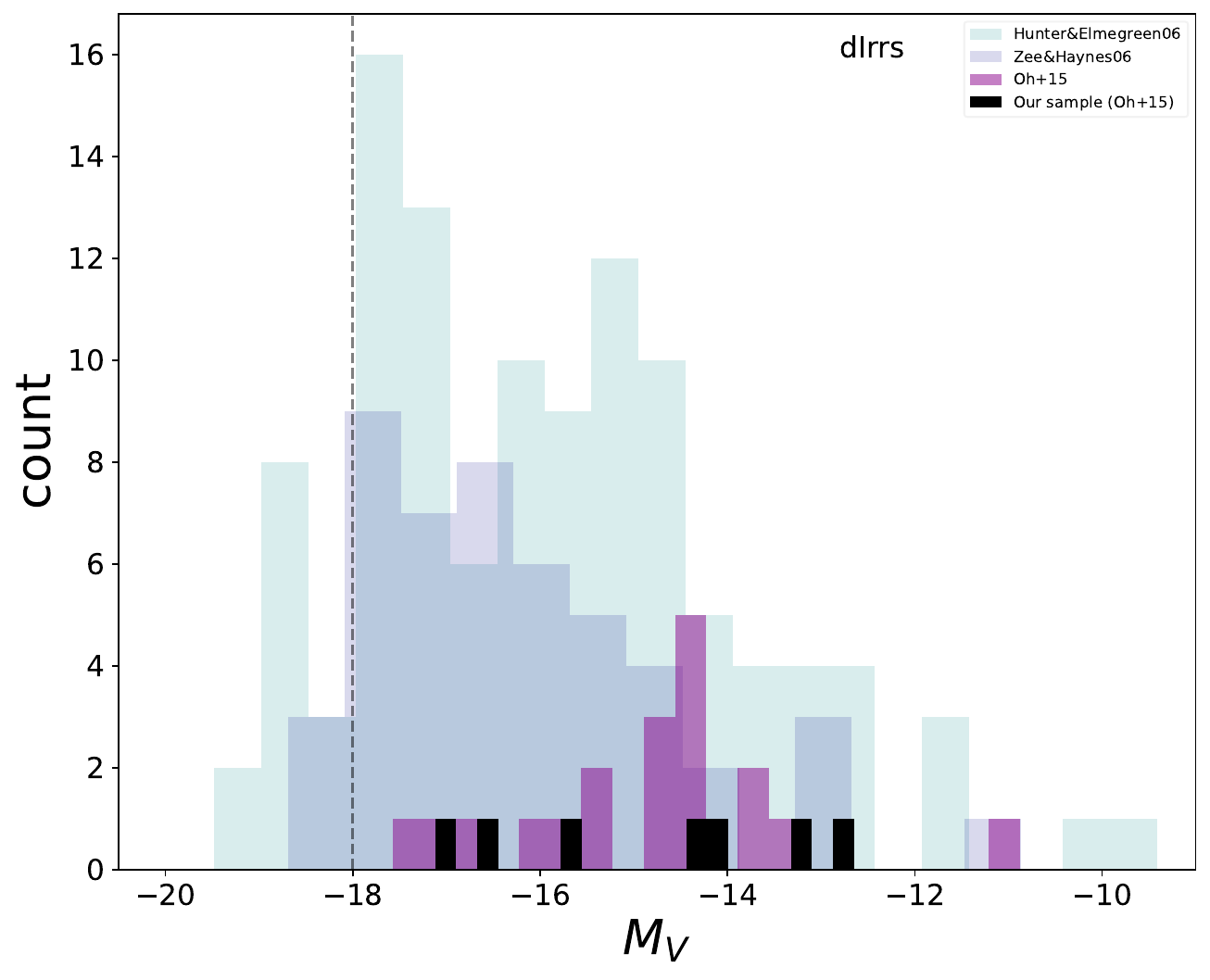}
\endminipage
\caption{Distribution of the general populations of [left] UDGs, [middle] LSBs and [right] dIrrs in the space of their defining physical parameters: $\mu_{g,0}-R_e$, $\mu_{B,0}$ and $M_V$, respectively. In each case, we denote the sample galaxies used for constructing dynamical models in black. The dotted lines indicate the ranges of respective parameters allowed for these galaxy classes.}
\label{pic:range}
\end{figure*}
\begin{figure}
    \centering
    \includegraphics[width=0.8\linewidth]{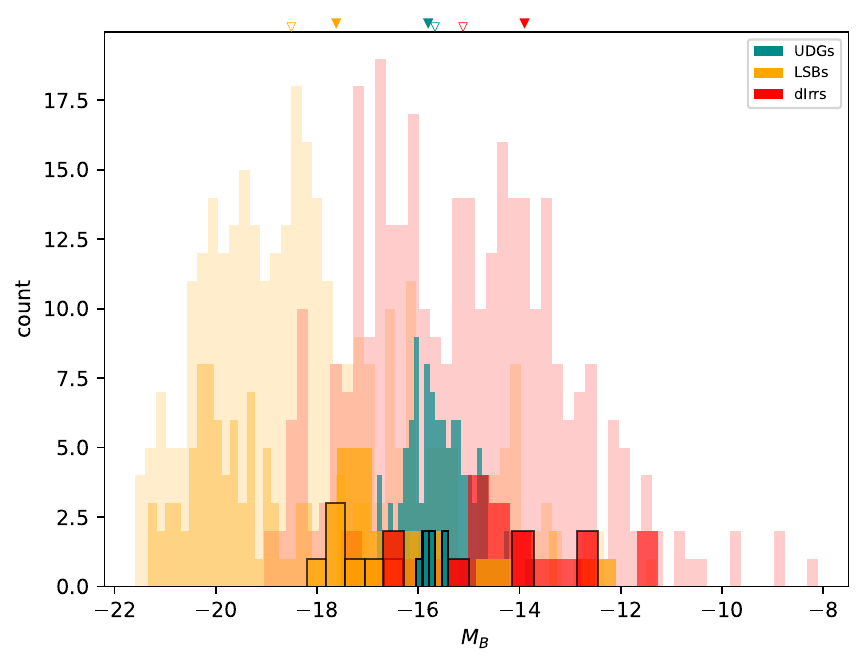}
    \caption{Distribution of the UDGs, LSBs and dIrrs in the $M_B$ space with UDGs, LSBs and dIrrs being represented by, respectively, teal, yellow and red colours. The empty and solid triangles on top of the plot represent the median $M_B$, respectively, of the parent populations and the samples considered in our work. The galaxies are chosen from the literature as follows (ordered as the increasing shade of the histograms) : (1) UDGs : \cite{2017Leisalfalfa}; (2) LSBs  : \cite{1996impeyLSBsother}, \cite{2004KniazevLSB} and \cite{de1996h}; (3) dIrrs : \cite{2006hunterDirr} and \cite{2006dirrVanZee}. The histograms with black outlines represents the samples considered in our study.}
    \label{fig:Mb}
\end{figure}
\section{Sample of galaxies}\label{sec:sample}
    \noindent For the scaling relations, we consider the following samples: (i) 7 UDGs from \cite{mancera2020robust} and \cite{kong2022odd}, (ii) 19 LSBs from \cite{de1996h} and \cite{2001deBlokMassmodelLSB}, (iii) 26 dIrrs from the \texttt{LITTLE THINGS} survey \citep{oh2015high} based on the availability of requisite data. 
    For dynamical models, the galaxy samples considered are as follows: (i) 7 UDGs from \cite{mancera2020robust}, \cite{shi2021cuspy}, \cite{kong2022odd}, (ii) 7 LSBs from \cite{de1996h}, (iii) 7 dIrrs from  \cite{oh2015high}, based on the availability of the mass modelling data in the literature as well as the quality of the modelling \textcolor{black}{(discussed in the next section)}. We may note that the galaxy AGC114905 of our UDG sample is extensively studied by \cite{2022PinaNoDM,2024PinaNoDM}. \\ \\
    \noindent To look for possible selection effects, we next compare each of our galaxy samples with a general sample collected from different sources. In Figure \ref{pic:range}(a), we present the distribution of all isolated, HI-rich UDGs as detected in the ALFALFA survey in the $\mu_{g,0}$-$R_{e}$ space \citep{2017Leisalfalfa}.  In Figure \ref{pic:range}(b), we show the distribution of the $\mu_{B,0}$ values of LSBs from the catalogue of \cite{1996impeyLSBsother}, and also from \cite{de1996h}, \cite{2004KniazevLSB}. 
    \begin{table}[h!]
        \centering
        \caption{The definitions of the UDGs, LSBs and the dIrrs.}
        \begin{tabular}{c c c c}
            \hline
            Galaxy class & Parameter & & Definition \\
            \hline
            \hline
            UDG & $\mu_{g,0}$, $R_e$ & & $>$ 24 mag arcsec$^{-2}$, $>$ 1.5 kpc\\
            LSB & $\mu_{B,0}$ & & $>$ 23 mag arcsec$^{-2}$\\
            dIrr & $M_V$ & & $>$ -18 \\
            \hline
        \end{tabular}
        \label{tab:definition}
    \end{table}
    In Figure \ref{pic:range}(c), we present the distribution of $M_V$ values of dIrrs from \cite{2006hunterDirr}, \cite{2006dirrVanZee} and \cite{oh2015high}. Table \ref{tab:definition} presents the values of different physical parameters for each galaxy type on which their respective definitions are based on (also, see Figure \ref{pic:range}).
    In Figure \ref{pic:range}(a), the black data points represent the UDGs we considered for constructing both scaling relations and dynamical models. 
    In Figure \ref{pic:range}(b) (\ref{pic:range}(c)), the black data points indicate the LSBs (dIrrs) considered for building dynamical models; the scaling relations, on the other hand, are based on both the black and the purple data points. We  note, our sample galaxies span fairly well the range of values corresponding to their defining properties, as exhibited by their general populations. Thus we may consider our galaxy samples to represent their respective parent populations well and relatively free of selection effect. However, our sample size is small and limited due to the prior availability of mass models in the literature. A larger sample size, if and when available, will help fine-tune results. \\
    
    \noindent We have followed the widely-accepted definitions for the different galaxy types studied in this work, namely the UDGs, the LSBs and the dIrrs. It is true that the definitions of the different galaxy types are based on different physical parameters, namely surface brightness and stellar disc scale length for the UDGs, absolute magnitude for the dIrrs and surface brightness for the LSBs. 
    So, it may appear that one galaxy can be both an LSB and a dIrr at one time. Therefore, we now compare and contrast the distribution of the parent populations of the above galaxy types with respect to their $B$-band absolute magnitude. in Figure \ref{fig:Mb}, we show the distribution of $B$-band absolute magnitude $M_B$ for all galaxies to study the possible overlap between different galaxy classes. The teal, yellow and red histograms combine all the catalogues for UDGs, LSBs and dIrrs respectively. 
    Although we do notice an overlap of the distributions, interestingly, the median values, as indicated by the empty triangles on the top, are fairly different in each case. In fact, the median $B$-band absolute magnitude of the LSB is starkly different from that of the dIrrs and the UDGs. Therefore, this indicates that a prototypical dIrr can never be a prototypical LSB and so on. Similarly, we should have compared the dIrrs and the LSBs with respect to the $B$-band central surface brightness distribution as well. 
    However, information on the $B$-band central surface brightness is not available for the dIrr population. This is because the catalogues mostly provide information on the physical parameters on which the definition of the galaxy type is based. The full width at half maxima of the HI spectra (W$_{50}$) serves as a proxy for the dynamical mass. For LSBs, it is 200 km s$^{-1}$ (\citealt{2001MathewsLSB}) and for dwarfs, it is 100 km s$^{-1}$ (\citealt{2001KarachentsevdIrr}). 
    However, we may note here that the median W$_{50}$ value for the same of LSBs from \cite{de1996h} is about 120 km s$^{-1}$. Therefore, it is reasonable to assume that UDGs, dIrrs and LSBs constitute three different types of galaxies.
    \\ \\
    \noindent Finally, since the number of galaxies is not large enough to carry out a statistical analysis like the PCA, we create mock galaxies corresponding to each galaxy in the samples for constructing dynamical models. We use the Monte Carlo (MC) method for the same (see, for example, \citealt{raychaudhuri2008introduction}). 
    Considering the distribution of values of the different input parameters, 300 mock galaxies are generated from each original galaxy in a sample. Thus we have an enlarged sample of 2,100 galaxies for each of the UDG, LSB and dIrr populations. Further, we construct dynamical models for each of these 2,100 galaxies in each of our galaxy samples. The model kinematic parameters thus obtained along with other structural parameters available in the literature are used to carry out PCA.

\section{Input parameters}\label{sec:input}
    \noindent We construct the dynamical models of the original and mock galaxies using input as well as trial parameters from the mass models available in the literature. In the AGAMA framework, for generating the net potential of the stars, gas and the dark matter halo, input parameters namely (i) the dark matter core density ($\rho_0$) and core radius ($R_c$)  (ii) the gas central surface density ($\Sigma_{HI}$), scale radius ($R_{D,HI}$) \& scale height ($h_{z, HI}$), as well as trial parameters for the \textcolor{black}{DF of the} stellar component,
    namely  (iii) stellar central surface density ($\Sigma_0$), exponential scale radius ($R_{D}$) \& scale height ($h_{z,*}$) as well as \textcolor{black}{stellar} radial ($\sigma_r$) and vertical ($\sigma_z$) velocity dispersions are required. 
    \\
    
    \noindent For the UDGs, the dark matter halo and the gas disk parameters are taken from \cite{kong2022odd}; $\Sigma_{HI}$ and $R_{D,HI}$ are obtained by extracting data from the gas density plots presented in \cite{kong2022odd} using HTML-based software WebPlotDigitizer\footnote{\cite{WebPlotDigitizer}} and then fitting a sech-squared profile using python module scipy.
    To set the dark matter halo potential for UDGs, the $\alpha$, $\beta$ and $\gamma$ values of Equation \ref{eq:111} are set according to the values listed in Table 2 of \cite{kong2022odd}.
    For the stellar component, we use the relation $M_* $ = $ 2\pi\Sigma_0R_D^2$ to calculate the stellar surface density of the galaxies, assuming their stellar disk component to be exponential in nature \citep{freeman1970disk}, $M_*$ and $R_D$ already presented by \cite{mancera2020robust}. 
    The central stellar surface density and stellar scale radius of the AGC242019, however, is obtained by fitting $\Sigma_* = \Sigma_0 \exp({-R/R_D})$ to the surface density profile presented by \cite{shi2021cuspy}. Similarly, for the LSBs, the dark matter halo parameters are taken from \cite{2001deBlokMassmodelLSB} and the gas disk parameters from \cite{de1996h}. 
    The central surface brightness values in B-band ($\mu_{B,0}$) and stellar scale radii are adopted from \cite{de1996h}. The stellar surface brightness is converted to stellar surface density using the $M/L$ ratio for $B-$band as presented in \cite{2001bellMbyL}. 
    Finally, for the dIrrs, we obtained the trial parameters of the stellar and gas components by extracting the data from the stellar and gas surface density profiles presented in \cite{oh2015high} and dark matter parameters from their Table 2. 
    All these parameters for our sample galaxies are listed in Table \ref{table:Properties}. 
    \\ \\
    \noindent The stellar scale height $h_z$ is one of the trial parameters which is not observationally measured for our sample galaxies. However, for ordinary edge-on galaxies, $h_z$ has been found to lie between $R_D$/5 and $R_D$/7 \citep{kruit1981hz}. 
    So we choose a typical value of $h_z$ = $R_D$/6 for our galaxy samples. 
    Interestingly, we note there is hardly any variation in the profiles with the change in the assumed values of $h_z$  (see Sect. \ref{sub:agama}, Figure \ref{pic:agama}).  
    This is because dispersion $\sigma$ varies as $h_z^{0.5}$ in a self-gravitating system of stars in dynamical equilibrium, and so the dependence of $\sigma$ on the assumed value of $h_z$ may be understood to be weak, in general. Besides, the galaxy models are obtained by iterative self-consistent method as described in Sect. 
    \ref{sec:model}, and thus the trial $h_z$ value can be considered as an initial guess for constructing the potential. Therefore, the error in the results due to the uncertainty in the $h_z$ values is also negligible. 
    \\
    
    \noindent Dynamical models for a sample of 5 dwarf galaxies by \cite{2011BanerjeeGasHz} showed that the gas scale heights $h_{z,HI}$ at a mean radius (1.5$R_D$) range between 0.2 - 0.4 kpc (0.1 - 0.4$R_D$). Besides, Figure 7 of \cite{2022ManceraImpact} indicates that the mean gas scale height of low-mass galaxies is about ranges between 0.2 - 0.6 kpc, which is similar to the range of values theoretically determined by \cite{2011BanerjeeGasHz}. So, we assume a $h_{z,HI}$ of 0.25$R_D$ for our sample galaxies. We further check and confirm that our results hardly change if we vary the value of the assumed $h_{z,HI}$ between 0.1 - 0.4 $R_D$. \\
    
    \noindent \textcolor{black}{Finally, a trial value is used for $\sigma_z$ in the DF based on the single-component model of a self-gravitating stellar disc, $\sigma_z = \sqrt{\pi G h \Sigma_0}$ (see for example \cite{mancera2020robust}) with h $\sim$ $R_D$/6 (as discussed above). Similarly, $\sigma_r$ is considered to be $\sigma_z/0.3$ \citep{2012dispGerssen}. }
    \noindent \textcolor{black}{By trial-and-error method, we fine-tune the value of $\sigma_z$ such that the output $h_z$ from the self-consistent model of AGAMA lies between $R_D$/5 $>$ $h_z$ $>$ $R_D$/7 (in addition to complying with the observed total rotation curve and stellar surface density profiles).}
    \\
    
    \noindent \textcolor{black}{We choose only a subset of the LSBs and the dIrrs studied in scaling relation to construct the dynamical models. The first selection criterion is based on the availability of the requisite stellar surface density and mass modelling data and the quality of fits of the exponential/sech-squared function to the surface density profiles to recover the input parameters for AGAMA. Further, we note a decrease in the stellar central surface densities more than 30\% for some galaxies while constructing the dynamical models (see Sect. \ref{sec:model}).}

\section{Methodology}\label{sec:method}
\subsection{Creation of mock galaxies}\label{sec:mock} 
    \noindent The mock galaxies generated from a parent galaxy represent a collection of galaxies with their physical parameters differing from those of the parent galaxy within the error bars.  We choose one parent galaxy of our sample (Table \ref{table:Properties}) and consider each of the ten input plus trial parameters \{$p_1$, $p_2$, $\dots$, $p_{10}$\} of the galaxy (Sect. \ref{sec:input}) to have a Gaussian distribution, with mean equal to the value corresponding to the parent galaxy and standard deviation equal to its error.  
    From the multi-Gaussian distribution of \{$p_1$, $p_2$, $\dots$, $p_{10}$\} thus obtained, we create a mock galaxy by randomly choosing ten parameters \{$p'_1$, $p'_2$, $\dots$, $p'_{10}$\} by employing an MC simulation. Thus 300 mock galaxies are generated from each of the parent galaxies. The dynamical models of these mock galaxies are constructed using AGAMA, the output of which is used in the PCA.
\\

\subsection{Principal Component Analysis}\label{sec:pca}
    \noindent  Principal component analysis (PCA) is used to understand the relative importance of the parameters in accounting for the variance in a given dataset (for reference \citealt{bro2014principal}). In PCA, the covariance matrix in the chosen parameter space is solved whose $i$th diagonal element describes the covariance of the $i$th parameter $\mathrm{x_i}$ with itself i.e. the variance ($\rm{cov}(x_i,x_i)$), and the $ij$th component gives the covariance between $x_i$ and $x_j$ ($\rm{cov}(x_i,x_j)$), their respective expressions being :
    \begin{eqnarray}
        \rm{cov}(x_i,x_i) &=& \frac{\sum_{i = 1}^n (x_i - \Bar{x}_i)(x_i - \Bar{x}_i)}{(n - 1)}    \\
        \rm{cov}(x_i,x_j) &=& \frac{\sum_{i,j = 1}^n (x_i - \Bar{x}_i)(x_j - \Bar{x}_j)}{(n - 1)} 
    \end{eqnarray}
    In these expressions, each barred quantity denotes mean value of the corresponding parameter and $n$ is the number of data points taken into account. 
    The eigenvectors of this matrix are the principal components while their eigenvalues indicate the relative importance of the principal components. In principle, the data will cluster around the principal component with the largest eigenvalue in the principal component space, followed by the principal component with the second largest eigenvalue. Thus if we project the data from original space to the principal component space, we see variation in the data is mainly explained by the first two principal components with the largest eigenvalues, thus leading to a dimensionality reduction. The principal components can be considered as linear combinations of the original variables, and the coefficients of their variables are called loadings. Variable corresponding to the largest loading primarily describes variance in the data. \\

    \noindent To obtain the PCA loading values we follow these steps: (1) perform PCA analysis with the largest possible set of the structural and kinematical parameters for each galaxy class (2) take the vector-sum of the first two principal components with the largest eigenvalues (3) identify the parameter which has largest loading value and normalize other loadings by the same (4) eliminate the parameters with negligible loading values and repeat the procedure until the results converge.  In all the cases, we see the first two principal components explain nearly 80-90\% variance in the data. Finally, we end up with the most relevant set of structural and kinematic parameters to study the dynamics of our galaxy samples.  We employ the sklearn  module in python to perform the principal component analysis.


\section{Results and discussion}\label{sec:result}
    \noindent We present the results in three subsections. In Sect. 7.1, we look for correlations between pairs of basic structural properties for our LSB and dIrr populations, and check if our sample UDGs comply with either of them. In the Sect. 7.2, we construct dynamical models of our sample UDGs, LSBs and dIrrs and compare their model-predicted stellar kinematics. Finally, in Sect. 7.3, we carry out a PCA analysis for a set of structural and kinematic properties of each of our galaxy populations to identify the crucial parameters explaining the variance in their respective datasets.
\begin{figure*}         
\hspace*{.45cm}
\includegraphics[width=0.8\paperwidth]{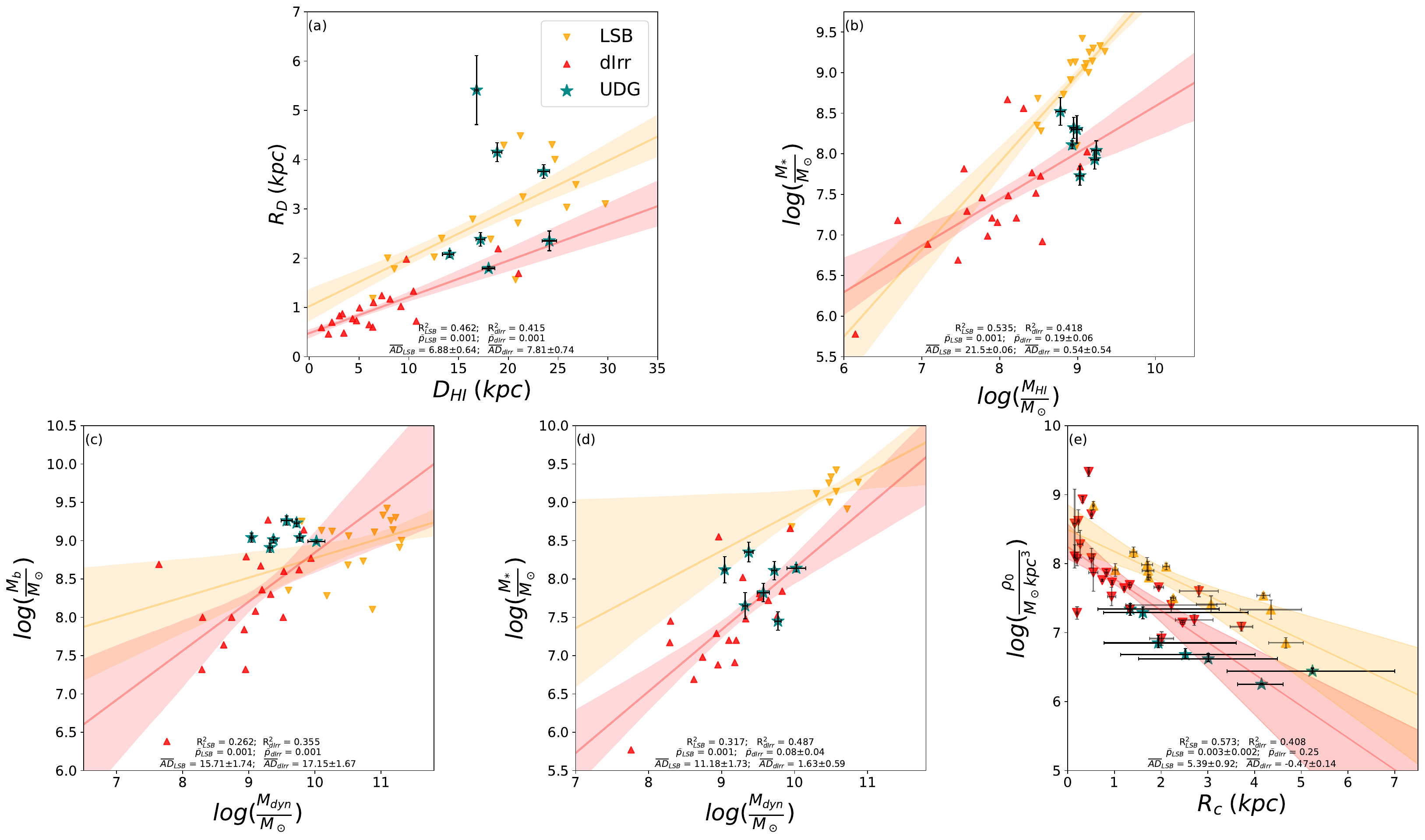}
\caption{Scaling relations between several mass and structural properties of our galaxy samples. 
In the top row, 
(a) HI-diameter versus exponential stellar scale radius ($D_{HI}$ vs $R_D$), 
(b) stellar mass versus HI mass (log $M_*$ vs log $M_{HI}$); 
in the bottom row, 
(c) total baryonic mass versus total dynamical mass enclosed in within 4R$_D$ (log $M_{b}$ vs log $M_{dyn}$), 
(d) stellar mass versus total dynamical mass (log $M_{*}$ vs log $M_{dyn}$), 
(e) dark matter core density versus core density ($\rho_0$ vs $R_c$). 
In all plots, the regression fits for LSBs and dIrrs with their 68\% (1$\sigma$) confidence interval  are shown in, yellow and red, respectively, and the data for UDGs are superposed on their regression fits with teal stars. The R$^2$-values for each of the regression fits for LSBs and dIrrs (R$^2_{LSB}$ and R$^2_{dIrr}$, respectively) are mentioned on the bottom of each plot. Also, $\bar{p}$ and $\overline{AD}$ are also mentioned on the bottom of each plot which measure the compliance of the UDG data with LSB or dIrr regression fit ($\bar{p}_{LSB}$, $\bar{p}_{dIrr}$, $\overline{AD}_{LSB}$ and $\overline{AD}_{dIrr}$, respectively) obtained from the two sample Anderson-Darling test. The steps of obtaining the $\bar{p}$ and $\overline{AD}$-values are mentioned in Sect. \ref{sub:pre}.}
\label{pic:input}
\end{figure*}
\subsection{Possible galaxy scaling relations}\label{sub:pre}
    \noindent In Figure \ref{pic:input}, we compare the results of statistical analysis of the physical properties of UDGs, LSBs and dIrrs. 
    \noindent For each pair of parameters, we determine the respective regression line fits with the 68\% ($\sim$ 1-standard deviation) confidence intervals for our sample LSBs and dIrrs, and superpose the data for our UDGs on the same. The error bars for the UDG data points are obtained from \cite{mancera2020robust} and \cite{kong2022odd} or evaluated by propagation of error. 
    We do not consider the error bars for most of the regression fits as these are not available for most of these parameters. We note the R-squared (R$^2$) value to measure the quality of the regression fit and reject the scaling relation for which the R$^2$ $<$ 0.25.
    We used python module statsmodels to obtain the scaling relations, fitting parameters and R$^2$s. We obtain the standard error in R$^2$ from a set of R$^2$s, each obtained by fitting regression to a simulated dataset generated from the original one, by Bootstrapping with replacement.
   Further, to check whether the UDGs comply with the regression fit of the LSB or the dIrr population, we do the following statistical test. For the pair of parameters (X,Y) involved in each of the scaling relations, we randomly choose 100 values of X in the same range as the x values of the UDG population. For each value of X so chosen, we generate the corresponding Y value by randomly choosing it from a Gaussian distribution centred at the Y value from the regression fit and standard deviation equal to half of the 68\% confidence interval from the same. Thus we generate a simulated dataset of 100 elements for each of the LSB and the dIrr populations. 
   Next, we conduct a two-sample Anderson-Darling (AD) test between the simulated LSB (dIrr) sample and the UDG sample, and note the $p$-value and the AD-coefficient. We repeat this process 10,000 times and note the mean $p$-value and AD-coefficient ($\bar{p}_{LSB}$, $\bar{p}_{dIrr}$, $\overline{AD}_{LSB}$ and $\overline{AD}_{dIrrs}$ respectively). The standard error in $p$ is mentioned unless negligible. The $\bar{p}$-value indicates the probability of the null hypothesis that the two samples are drawn from the same population. 
   A $\bar{p}$-value < 0.02 indicates that the two populations are significantly different from each other. On the other hand, the alternative hypothesis that the two samples are drawn from the same normal distribution cannot be ruled out for $\bar{p}$ $>$ 0.02. 
   Additionally, we compare $\overline{AD}$ with the critical $AD$-coefficient ($AD_{crit}$) for the sample sizes of the two populations considered in the AD test \citep{1976Pettitt_stat}. In all the cases, our $AD_{crit}$ has a value of 2.718 at a significance level of 2.5 per cent. For $\overline{AD}$ > $AD_{crit}$, the null hypothesis is rejected. We use the scipy module in python to perform these tests.
    \\
    
    \noindent We study the exponential scale radii ($R_D$) versus HI-diameters ($D_{HI}$) for our sample UDGs, LSBs and dIrrs in panel ($a$). It is well known that similar to dIrrs and LSBs, field UDGs have remarkably high HI-masses for their stellar content (\citealt{2017trujiliohimassofudg,mancera2020robust}). 
    For normal spiral galaxies, the ratio of the HI-to-optical diameter is nearly 2 \citep{1997broelisrhee}, yet there are several galaxies which have unusually extended HI-disks (\citealt{2005begumgiantHI}; \citealt{2007gentileExtended}; \citealt{2016Wang}). The extended HI disk can be attributed to gas accretion or merger, though the origin of such large gas disks is still debatable \citep{2005NoordermeerHI}. 
    The HI-mass ($M_{HI}$) and diameter ($D_{HI}$) of a galaxy has been found to be correlated as follows: log $D_{HI}$ = 0.506 log$M_{HI}$ - 3.293 (\citealt{1997broelisrhee,2016Wang}). UDGs in our galaxy sample except AGC242019 follow the D$_{HI}$-M$_{HI}$ relation of \cite{2016Wang} \citep{2021GaultVLA}. We use this relation to obtain the $D_{HI}$ of the galaxies and obtain the $R_D$-$D_{HI}$ scaling relation. The R$^2$ for the regression fits for LSBs and dIrrs are respectively, 0.462 and 0.415. In this context, we mention that for all the regression fits discussed in this section, the standard error of R$^2$ is negligible ($\sim$ 10$^{-5}$). 
    We next do an AD test to check if the UDGs comply with the scaling relation of the dIrr or the LSB. Both $\bar{p}_{LSB}$ and $\bar{p}_{dIrr}$ have a value 0.001 for this scaling relation meaning that the UDGs do not follow any of their normal distribution as far as the $R_D$-$D_{HI}$ scaling relation is concerned. A similar conclusion can be drawn by observing that both $\overline{AD}_{LSB}$ (6.88$\pm$0.64) and $\overline{AD}_{dIrr}$ (7.81$\pm$0.74) are greater than $AD_{crit}$.
    \\

     \noindent In panel ($b$), we present the correlation between the stellar mass ($M_*$) and the HI-mass ($M_{HI}$) of LSBs and dIrrs for which the R$^2$ values are, respectively, 0.535 and 0.418. We note that $\bar{p}_{LSB}$ and $\overline{AD}_{LSB}$ are again 0.001 and 21.5$\pm$0.06 for this scaling relation thus rejecting the null hypothesis. However, $\bar{p}_{dIrr}$ is 0.19$\pm$0.06 and $\overline{AD}_{dIrr}$ is 0.54$\pm$0.54 suggesting that the UDGs and dIrrs may follow similar normal distribution. Although the range of HI masses of the UDGs is the same as the LSBs, they, however, obey the same $M_*$-$M_{HI}$ scaling relation as the dIrrs. Since star formation rate increases with increasing stellar mass, UDGs and dIrrs will have relatively smaller star formation rate than the LSBs (Figure 8 of \citealt{2018zhouHImassStellarmass}). Hence, the extraordinary low surface brightness of UDGs may be attributed to inefficient star formation in these galaxies (as suggested by \cite{1997mcgaugh}, \cite{1997impeyReviewLSB} for LSBs). 
\\

    \noindent Since dwarf galaxies are dark matter dominated at all radii, dark matter plays an important role in their dynamics and can be crucial to explain their low luminosity (\citealt{2016Read,2017Read,2022PinaNoDM}).
    \cite{2017GargBanerjee} showed that dark matter suppresses the local axisymmetric instabilities which, in turn, may lower the star formation rate in a galaxy leading to a lower surface brightness. Hence, in the $bottom$ row of Figure \ref{pic:input}, we discuss the correlations between dynamical parameters that may explain the role of dark matter in regulating their evolution.
\\

    \noindent In panel ($c$), we present the respective regression fits between baryonic ($M_{b}$) and dynamical ($M_{dyn}$) mass with the 1$\sigma$-confidence intervals, for LSBs and dIrrs. The dynamical mass is taken to be the mass enclosed  within the radius of 4 times the stellar scale radius $R_D$ ($=4 V^2_\infty R_D/G$) , the baryonic mass being $M_* + 1.33M_{HI}$ (1.33 being the helium correction factor). This is analogous to the baryonic Tully-Fisher relation (BTFR) where the dynamical mass is be considered as a proxy of their circular velocity (\citealt{2000McGaughetal,2005McGaugh}). The R$^2$ for the regression fits for LSBs and dIrrs are respectively, 0.262 and 0.355. Similar to panel ($a$), ($\bar{p}_{LSB}$, $\bar{p}_{dIrr}$) and ($\overline{AD}_{LSB}$, $\overline{AD}_{dIrr}$) have values (0.001, 0.001) and (15.71$\pm$1.74, 17.15$\pm$1.67), respectively, thus suggesting that the UDGs constitute a distinct class of galaxies, relatively baryon-dominated, lying above the BTFR as discussed in \cite{2019pinaoffBTFR}.
    \\ 
\\ 

    \noindent In panel ($d$), we show the correlation between stellar mass ($M_*$) and dynamical mass ($M_{dyn}$) for our galaxy samples. The R$^2$ values for the regressions fits of LSBs and dIrrs are, respectively, 0.317 and 0.487. We note that the value of $\bar{p}_{LSB}$ = 0.001 and $\overline{AD}_{LSB}$ = 11.18$\pm$1.73 from which we may infer the UDGs and LSBs are members of two different populations.
    However, $\bar{p}_{dIrr}$ is 0.08$\pm$0.04 and $\overline{AD}_{dIrr}$ is -1.63$\pm$0.59 thus we cannot decline the possibility that the UDGs and the dIrrs may belong to a similar galaxy population. This is in line with the fact that the dIrrs share similar stellar and dynamical mass ranges, with the UDGs having slightly higher values of the same.
    In a way $M_*$ and $M_{dyn}$ scaling relation indicates the star formation rate as a function of the dark matter mass of a galaxy. Hence, the difference in their correlation may imply the difference in the way the dark matter regulates the star formation rate. \\ \\
    Finally, in panel ($e$), we show the correlation between dark matter core density ($\rho_0$) and core radius ($R_c$) which is routinely used to study the dark matter dominance in galaxies (\citealt{2001deBlokMassmodelLSB,2017bapatbanerjee}). The error bars for the halo core radius are obtained from \cite{kong2022odd}, \cite{2001deBlokMassmodelLSB} and \cite{oh2015high}, respectively, for the UDGs, LSBs and dIrrs. We obtained the weighted regression fits for the LSBs and dIrrs using the python module statsmodels and the corresponding R$^2$ equal to 0.573 and 0.408, respectively. Based of the values of $\bar{p}_{LSB}$ (= 0.003$\pm$0.002) and $\overline{AD}_{LSB}$ (= 5.39$\pm$0.92 $>$ $AD_{crit}$), we conclude that the LSBs and UDGs are very different from each other with respect to their dark matter distributions. However, $\bar{p}_{dIrr}$ (= 0.25) and $\overline{AD}_{dIrr}$ (= -0.47$\pm$0.14 $<$ $AD_{crit}$) indicate that the UDGs and the dIrrs may follow the same normal distribution and hence evolve under similar dark matter halo potentials. Further, we observe, for a given $\rho_0$ value, the LSBs have more cuspy halos, while compared to the UDGs and the dIrrs.
\\ \\
    \noindent Considering the comparison of all the scaling relations above, we may say that UDGs constitute a different population as far as the LSBs are concerned. However, one cannot completely rule out the possibility of the UDGs and the dIrrs  being one and the same population.

\begin{figure*}
\includegraphics[width=1.\linewidth]{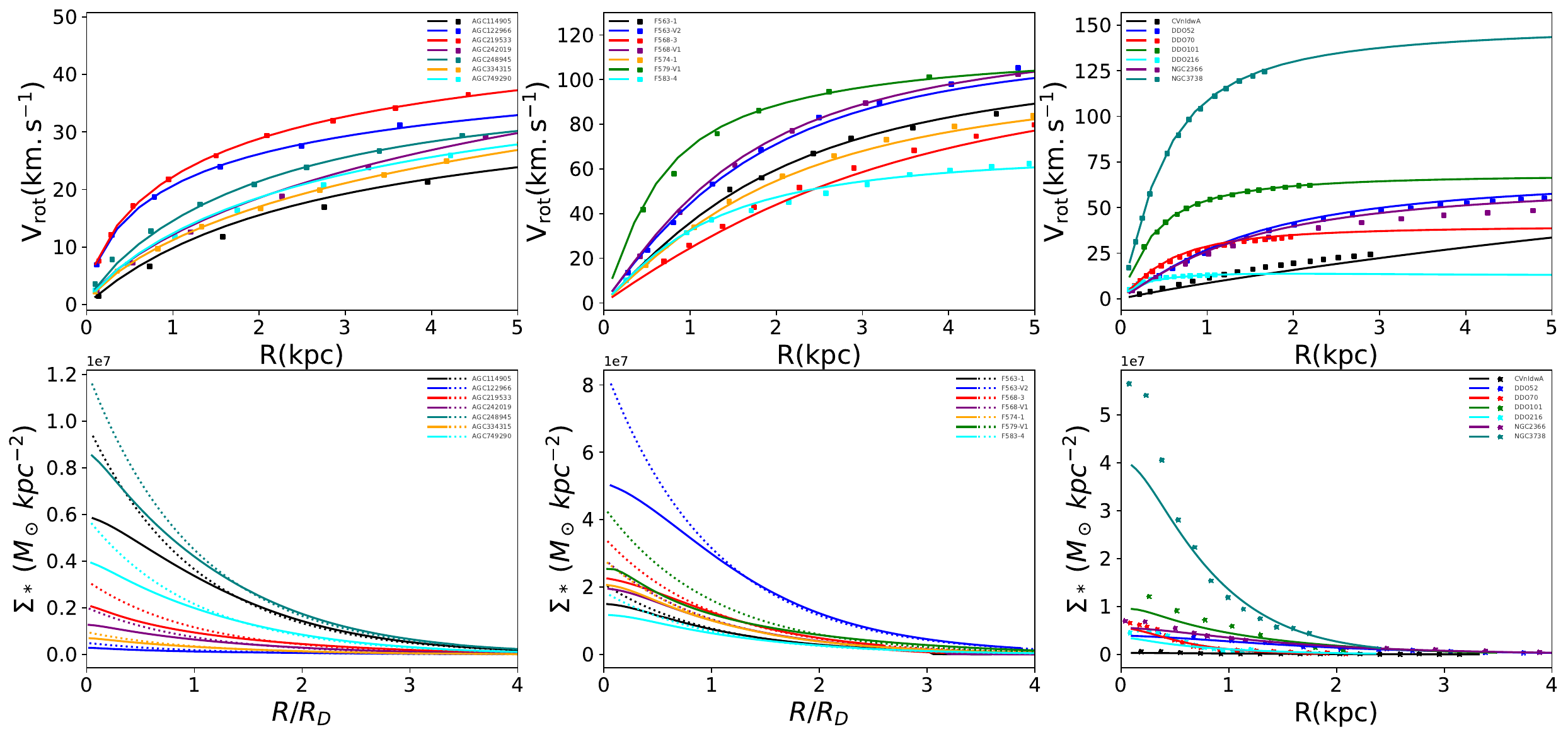}
\caption{Comparison between model (solid lines) and observed (scatter data points/dotted lines) (i) rotation curves (top panel) and (ii) stellar surface density profiles (bottom panel), for UDGs (left), LSBs (middle) and dIrrs (right).}
\label{pic:out}
\end{figure*}

\subsection{Stellar kinematics from dynamical modelling using AGAMA}\label{sub:agama}

    \noindent Stellar kinematics constitute a primary diagnostic tracer of the dynamical and secular evolution of a galaxy \citep{2018PinnaSVE}. We generate galaxy models for our UDG, LSB and dIrr samples using the input parameters listed in Table \ref{table:Properties} employing AGAMA. We constrain the galaxy models by the observed rotation curve and surface density profiles.
    To check the consistency between the galaxy models and the observed profiles, we compare and contrast the model-predicted rotation curves and stellar surface density profiles with the observed ones. In Figure \ref{pic:out}, we present the observed profile by points and the model-obtained profiles with solid lines. For the stellar surface density of the UDGs and LSBs, we reconstructed the exponential profile from their central surface density and stellar scale radius given in the literature. The same is shown with the dotted lines. We note that the central values of the stellar surface density is slightly smaller compared to their observed profile (see Sect. \ref{sec:model} for a discussion).
\\ \\

\begin{figure}
\hspace*{-0.5cm}
\includegraphics[width=1.\linewidth]{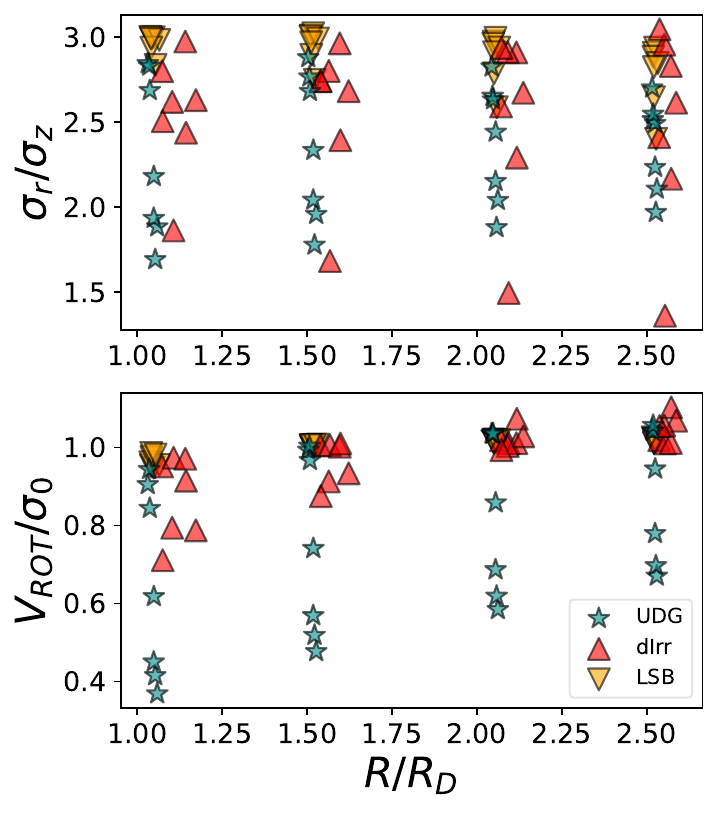}
\caption{Top panel : The ratios of stellar radial-to-vertical velocity dispersion ($\sigma_r/\sigma_z$), obtained from the AGAMA at $1R_D$, $1.5R_D$, $2R_D$ and $2.5R_D$ plotted for UDGs, LSBs and dIrrs. Bottom panel : Their circular-to-total stellar velocity dispersion ratios ($V_{ROT}/\sigma_0$)  at the same radii as the top panel.}
\label{pic:agama}
\end{figure}

    \noindent In the top panel of Figure \ref{pic:agama}, we plot stellar $\sigma_r/\sigma_z$ at a few galactocentric radii $R$ normalized by their exponential stellar disk scale length $R_D$, for our sample of LSBs, UDGs and dIrrs. 
    $\sigma_r/\sigma_z$ may indicate the relative dominance of stellar disc heating in the radial versus the vertical direction. 
    While $\sigma_z$ is mostly regulated by vertical bending instabilities (see for example, \citealt{2019lokasbuckling}), $\sigma_r$ is governed by radial heating due to disk non-axisymmetric features like bars and spiral arms (\citealt{2018PinnaSVE} and the citations therein). 
    According to \cite{2003shapiroobs}, late-type galaxies have $\sigma_z/\sigma_r$ $<$ 0.5 (i.e., $\sigma_r/\sigma_z$ $>$ 2), while early-type galaxies have $\sigma_r/\sigma_z < 2$. 
    We observe that at almost all radii, the $\sigma_r/\sigma_z$ values of the UDGs range between 1.5 - 3, and the corresponding values for the dIrrs seem to mostly overlap with them. While on the other hand, the average $\sigma_z/\sigma_r$ for the LSB population is relatively higher, lying between 2.5 - 3 consistently at all radii. 
    In addition, we conduct a statistical test to quantitatively assess the similarities between the kinematic parameters of two galaxy samples. For each galaxy sample, we resample the data of $\sigma_z/\sigma_r$ at a galactocentric radius by Bootstrapping with replacement to increase the sample size to 100. Next, we perform the non-parametric Mann-Whitney U-test between UDGs and LSBs (dIrrs) considering the $\sigma_z/\sigma_r$-values at one radius and note the corresponding p-value, $p_{LSB}$ ($p_{dIrr}$). We repeat the process 1,000 times at each of the 20 radial points between 0 to 4$R_D$ and then obtain the median $p$-value, $\Tilde{p}_{LSB}$ ($\Tilde{p}_{dIrr}$) at each radius. We note that $\Tilde{p}_{LSB}$ and $\Tilde{p}_{dIrr}$ values are of the order of 10$^{-26}$ and 10$^{-26}$  respectively that as far as the shape of the stellar velocity ellipsoids are concerned, UDGs, LSBs and dIrrs constitute different populations. \\

    \noindent In the $bottom$ panel of Figure \ref{pic:agama}, we present the ratio of the rotational velocity-to-the-total stellar velocity dispersion ($V_{ROT}/\sigma_0$). $V_{ROT}/\sigma_0$ is a measure of the angular momentum-to-random motion support against gravitational collapse in a galaxy (\citealt{2009tolobadwarfrotator,2022beyondgalaxy}). 
    \cite{2019FalconBorrosoCALIFA} found that elliptical galaxies have relatively lower values of $V_{ROT}/\sigma_0$ which increases as we move from left to right in the Hubble sequence. Low and high values of $V_{ROT}/\sigma_0$ indicate slow and fast rotators respectively. 
    We note that at the inner radii, $V_{ROT}/\sigma_0$ for the LSBs lie between 0.7 and 1, while at the intermediate and the outer radii, the value is close to 1, indicating that LSBs are fast rotators. This is in line with their late-type kinematics found  above. 
    For the dIrrs and the UDGs, the average value of $V_{ROT}/\sigma_0$ lies around 0.5 at the inner radii. At the intermediate and outer radii, the average $V_{ROT}/\sigma_0$ for the dIrrs are 0.8 and 0.9 respectively, though with a large variance. For the UDGs, the corresponding both the values are about 0.6, again with a fairly large scatter. 
    Further, we conduct the Mann-Whitney U-test and note the $\Tilde{p}_{LSB}$ and $\Tilde{p}_{dIrr}$ values. We observe that $\Tilde{p}_{LSB}$ and $\Tilde{p}_{dIrr}$ have values very close to 0 thus rejecting the hypothesis that the UDGs and LSBs (dIrrs) constitute the same population.

    \noindent Therefore, we conclude that UDGs, LSBs and dIrrs are very different as far as their model-predicted kinematics is concerned. Here, we emphasise on the fact that the values of $\sigma_r/\sigma_z$ and $V_{ROT}/\sigma_0$ presented in this section are obtained from their dynamical models.
    \\

%
%
%
%
%
%
\begin{figure}
\includegraphics[width=1.\linewidth]{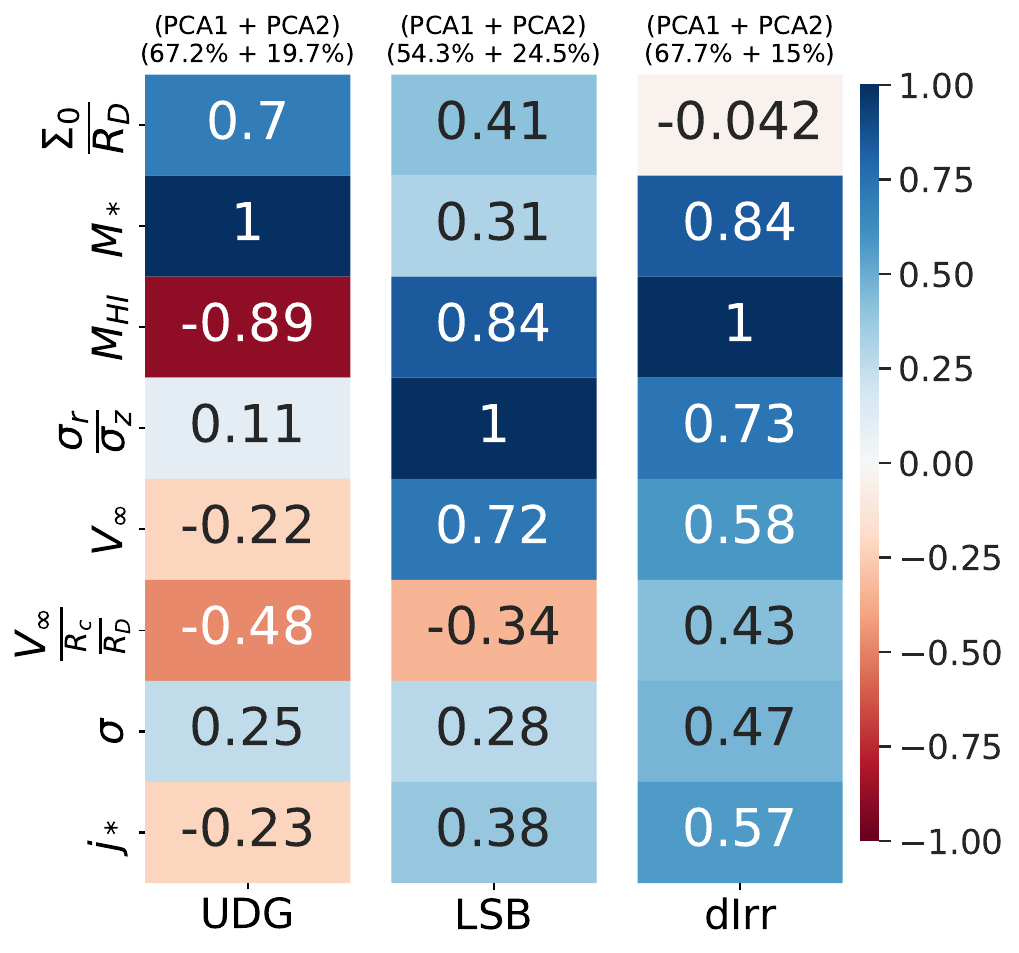}
\caption{Principal component analysis : Loadings of first two principal components  presented for each of our UDG, LSB and dIrr galaxy samples, which account for at least 80\% of the variation in the data. Parameters considered in this analysis from top to bottom are as follows : the ratio of stellar central surface density-to-exponential stellar scale radius ($\Sigma_0/R_D$), stellar mass ($M_*$), HI-mass ($M_{HI}$), ratio of stellar radial-to-vertical velocity dispersion ($\sigma_r/\sigma_z$), asymptotic velocity ($V_\infty$), compactness parameter ($V_\infty/(R_c/R_D)$), stellar velocity dispersion ($\sigma$) and specific angular momentum ($j_*$).}
\label{pic:load}
\end{figure}

\subsection{The key dynamical parameters of our galaxy samples: Principal Component Analysis}\label{sub:princ} 
    \noindent Finally, in Figure \ref{pic:load}, we present the results obtained from the PCA of a set of dynamical parameters possibly driving their disk dynamics for each of our galaxy samples. Through trial and error, we choose the following set of dynamical parameters for our galaxy samples (for details Sect. \ref{sec:pca}).
    \begin{itemize}
    \item Ratio of stellar surface density to scale radius ($\Sigma_0/R_D$) : By definition UDGs have very low central surface brightness with relatively large effective radii, compared to other low surface brightness galaxies \citep{2015vanforty}. Thus we may consider the ratio of stellar surface density to stellar scale radius as a possible defining parameter of UDGs. 
    \item Stellar mass ($M_*$) : The stars constitute one of three basic building blocks of a galaxy, the others being the gas and dark matter halo. The stellar mass correlates with the dark matter mass as well as with the star-formation histories thus revealing key information about the formation and evolution of the galaxy \citep{2009conroymstarsfr}.
    \item Atomic hydrogen mass ($M_{HI}$) : $M_{HI}$ correlates with several fundamental galaxy properties, thus serving as a fingerprint of the stage of evolution of a galaxy  \citep{1997mcgaughgasmass}. It was observed that the correlation between galaxy gas content and the central surface brightness of disk galaxies is the strongest, making $M_{HI}$ a crucial dynamical component especially for low-surface brightness galaxies.
    \item Asymptotic velocity ($V_\infty$) : $V_\infty$ has a tight correlation with the total baryonic mass of a galaxy through baryonic Tully-Fisher relation. Moreover, $V_\infty$ can be considered as a proxy of the total dynamical mass of a galaxy, and hence can be considered as a dynamical parameter, as understood seen in previous studies \citep{2023adityaHI}. We consider the value of mean velocity at a radius equal to $4R_D$ as the asymptotic velocity.
    \item Specific angular momentum ($j_*$) : Specific angular momentum is one of the fundamental dynamical parameters of a galaxy that traces the accretion history of the gas disk thus revealing important information about its formation and evolution. There is a strong correlation between the stellar specific angular momentum and the stellar mass of the galaxies followed by various morphological types observed over different redshifts (\citealt{1983fallrelation,2012romanfall}). In this connection, it may be noted that correlations between mass, stellar specific angular momentum and gas fraction/central surface density were observed by \cite{2021Pina_tightAM}, and \cite{2024Elson}.
    We have calculated the specific angular momentum as :
    \begin{equation}
        j_* = \frac{2\pi \bigintss_0^\infty \Sigma_*(R) v(R) RdR}{2\pi \Sigma_0 R^2}
    \end{equation}
    considering an exponential stellar surface density profile $\Sigma_*(R)$ and velocity profile to be : $v(R) = V_0 [1 - \exp({-R/a})]$ ($V_0$ and $a$ being the fitting parameters of the rotation curve).
    In our study, we calculate $j_*$ by considering the model-derived stellar surface density and rotation velocity profile. We perform the integration between 0 to $\infty$ using the python module scipy.
    \item Total stellar velocity dispersion ($\sigma$) : Rotation-supported self-gravitating disks are unstable against gravitational instabilities and random motion of stars quantified by velocity dispersion of the disk is crucial for their stability, as indicated by the Toomre-Q parameter \citep{toomre1964pressure}. Furthermore, the total stellar velocity dispersion shows a  correlation with the dark matter halo mass thus implying dark matter dominance in galaxy evolution \citep{2018zahidcorrelation}. AGAMA over-predicts the central surface density due to the caveat discussed in Sect. \ref{sec:model}. Thus, we consider the total stellar velocity dispersion value at 1.5R$_D$ where the model-obtained stellar surface density distribution matches the observed profile fairly well (see Figure \ref{pic:out}).
    \item Stellar radial-to-vertical velocity dispersion
    ($\frac{\sigma_r}{\sigma_z}$) : Stellar velocity ellipsoid (SVE) constitutes a fingerprint of the disk heating mechanisms of a galaxy. $\sigma_r/\sigma_z$ quantifies the flattening of the SVE and can be considered as an indicator of the secular evolution of a galaxy. \citep{1999kruitgrijs}. For PCA, we consider the value of $\sigma_r/\sigma_z$ at 1.5R$_D$ as it should be representative of the mean kinematics of the disk.
    \item Compactness parameter ($\frac{V_\infty}{\frac{R_c}{R_D}}$) : 
    The ratio of dark matter core radius-to-exponential stellar disc scale length ($R_c/R_D$) can be considered as an indicator of the compactness of the dark matter halo, with  $R_c/R_D \geq$ 2 and $R_c/R_D \leq$ 2 indicating non-compact and compact dark matter halo, respectively. (\citealt{2004GentileCompactnessHSB,2008Banerjeecompactness,2010Banerjeecompactness,2013BanerjeecompactnessLSB}). \cite{2023adityaHI} extended this definition further by introducing the ratio of asymptotic velocity ($V_\infty$) and $R_c/R_D$ as the compactness of mass distribution. The dynamical masses of low luminosity galaxies are primarily dominated by dark matter at all radii. Hence, $V_\infty/(R_c/R_D)$  indicates the compactness of mass distribution within the central region of the galaxies. Larger the value of $V_\infty/(R_c/R_D)$, more compact is the mass distribution. The compactness of mass distribution is argued to be an important parameter to understand the stability of superthin stellar disk (\citealt{2023adityaHI} and references therein).
    \end{itemize} 
    \noindent In Figure \ref{pic:load}, from left to right, we present the sum of the first two loadings of the PCA corresponding to our UDGs, LSBs and dIrrs, which account for $\sim$ 87\%, 79\%, and 83\% variance in the data respectively. We observe that, for the UDGs, $M_{*}$, $M_{HI}$ and then $\Sigma_0/R_D$ can explain the maximum variance in the data. For dIrrs, the important parameters are $M_{HI}$ and $M_{*}$, followed by $\sigma_r/\sigma_z$.
    As observed by \cite{mancera2020robust}, the mass fraction of dark matter in UDGs is relatively low, and therefore it is not surprising that $M_{HI}$ and $M_*$ have emerged as the fundamental dynamical parameters for the UDGs from the PCA analysis. We conclude, that for both the UDGs and the dIrrs, $M_{HI}$ and $M_*$ play key roles in regulating the disk dynamics as is evident from the PCA analysis of their dynamical parameters.
    However, $j_*$  also appears to be a crucial parameter for the  dIrrs but not for UDGs. This is in contradiction to the predictions of \cite{2016AmoriscoLoeb}, \cite{mancera2020robust} and \cite{2023Benavedis}, who argued that UDGs are possibly hosted in high-spin dark matter halo which may regulate their formation scenario. Curiously, this is in compliance with the findings by \cite{2017udgdirrBel} where they reported the discovery of two field dIrrs SdI-1 and SdI-2 which closely follow the properties of UDGs. Though their spectroscopic images reveal remarkable irregular features, their sizes are comparable to UDGs as demonstrated by their $V$-band magnitude-effective radii distribution and hence in further studies they are considered to be isolated UDGs (e.g. \citealt{2017Papastergis}). 
    For LSBs, on the other hand, $\sigma_r/\sigma_z$, $M_{HI}$ and $V_{\infty}$  mostly explain the variance in the data.

\section{Conclusions}\label{conclusion}
    \noindent In this work, we study the dynamical lineage of isolated, HI-rich UDGs and look for a possible common origin with other low luminosity galaxies, namely the LSBs and dIrrs. We consider a sample of galaxies for each of the above galaxy populations and first construct possible scaling relations between pairs of structural and kinematical parameters as obtained from the literature. We then obtain separate regression fits for the LSB and dIrr data with the corresponding 68\% confidence intervals, and superpose the data points for the UDGs on them. We further conduct two-sample Anderson-Darling test to check the similarities among UDGs, LSBs and dIrr samples. We observe that the UDGs and LSBs constitute statistically different populations. However, based on the stellar mass versus HI mass (log $M_*$ - log $M_{HI}$), stellar mass versus total dynamical mass (log $M_*$ - log $M_{dyn}$) and dark matter core density and core radius ($\rho_0$ - $R_c$) fits, we cannot rule out the possibility of the UDGs and the dIrr populations following same normal distributions.
    Next, we construct their dynamical models employing the publicly available stellar dynamical code AGAMA as constrained by their mass models already available in the literature. A Mann-Whitney U-test on the distribution of the kinematical parameters, namely ratios of radial-to-vertical stellar velocity dispersion and the rotational velocity-to-total stellar velocity dispersion, indicates that the UDGs, dIrrs and the LSBs constitute starkly different populations. Finally, we carry out a Principal Component Analysis of a fixed set of structural and kinematical parameters for each galaxy population which plausibly regulates their disk dynamics. We find that the variance in the data of the UDGs and the dIrrs are mostly explained by a common set of parameters namely, the total HI mass and the stellar mass. For the LSBs, the same is accounted for by the radial-to-vertical stellar velocity dispersion ratio, followed by the HI-mass. Therefore, considering above, isolated, HI-rich UDGs seem to share a common dynamical lineage with the dIrrs but not so much with the LSBs.

\begin{acknowledgements}
      We acknowledge that the project is funded by the Prime Minister's Research Fellowship with ID: 0902007. We thank the anonymous referee for his/her detailed, constructive suggestions which have improved the paper. We also thank Prof. Nissim Kanekar for useful discussions and suggestions.  
\end{acknowledgements}
\bibliographystyle{aa.bst}
\bibliography{bib}{}
\begin{appendix} 
\onecolumn
\section{Input parameters for the galaxy modelling}
The input parameters for constructing the dynamical models the galaxies are listed in Table \ref{table:Properties}.
\begin{table*}[h!]
\begin{center}
\hspace*{10cm}
\centering

\caption{Input parameters for constructing the dynamical models of UDGs, LSBs and dIrrs.}
\begin{tabular}{l c c c c c c c}
\hline
Galaxy  & $\mu_0$ & Stellar & Exponential  & HI surface & Sech-squared &  DM core & DM core
\\
Names  & or & surface  & stellar scale & density & HI-disk scale &  density  &   radius
\\
& $M$\tablefootmark{a}& density ($\Sigma_{0}$) & radius ($R_{D}$)  & ($\Sigma_{HI}$) & radius ($R_{D,HI}$) &  ($\rho_{0}$) &  ($R_{c}$) 
\\
& & $M_{\odot}/pc^{2}$ & kpc  & $M_{\odot}/pc^{2}$ & kpc & 10$^6$ $M_{\odot}/kpc^{3}$ & kpc \\
\hline    
\hline    
UDGs\tablefootmark{b} & $\mu_{g,0}$ & & & & & &\\
\hline
AGC114905& 23.62 & 9.92 & 1.79  & 4.84 & 10.18 & 1.77 & 4.15 \\
AGC122966 & 25.38 & 0.49 & 4.15  & 3.89 & 12.14 & 21.87 & 1.34 \\
AGC219533 & 24.07 & 3.16 & 2.35  & 5.74 & 17.76 & 19.48 & 1.61 \\
AGC242019 & - & 2.01 & 5.41 & 3.41 & 3.26 & 11.14 & 5.24 \\
AGC248945 & 23.32 & 12.18 & 2.08  & 5.21 & 7.54 & 7.07 & 1.94 \\
AGC334315 & 24.52 & 0.96 & 3.76  & 6.35 & 17.92 & 4.26 & 3.01 \\
AGC749290 & 24.66 & 5.87 & 2.38  & 5.33 & 10.1 & 4.78 & 2.52 \\

\hline
LSBs\tablefootmark{c} & $\mu_{B,0}$ & & & & & &\\
\hline
F563-1 & 23.63 & 20.94 & 3.24 & 4.62 & 11.59 & 79 & 1.72 \\
F563-V2 & 22.10 & 85.71 & 1.56 & 7.41 & 8.02 & 96.6 & 1.7 \\
F568-3 & 23.08 & 34.75 & 3.03 & 4.61 & 10.82 & 25.7 & 3.07 \\
F568-V1 & 23.30 & 28.38 & 2.38 & 3.97 & 11.4 &  146.1 & 1.41 \\
F574-1 & 23.31 & 28.12 & 3.49 & 2.96 & 12.26 & 63.1 & 1.74 \\
F579-V1 & 22.83 & 43.75 & 3.1 & 3.61 & 10.9 &  694.4 & 0.55 \\
F583-4  & 23.76 & 18.58 & 2.02 & 1.88 & 5.97 &  80.6 & 1.02 \\

\hline
dIrrs\tablefootmark{d} & $M_V$ & & &  & & & \\
\hline
CVnIdwA & -12.4 & 0.79 & 0.83  & 10.71 & 10.18 & 8.19 & 2.01 \\
DDO 52 &-15.4 & 6.26 & 1.33  & 5.94 & 4.55 & 48.81 & 1.33 \\
DDO70 & -14.1 & 8.58 & 0.48  & 7.94 & 0.97 & 119.95 & 0.51 \\
DDO101 & -15.0 & 16.61 & 0.87  & 3.19 & 2.2 & 849.14 & 0.32 \\
DDO216 & -13.7 & 6.48 & 0.59  & 2.16 & 0.88 & 127.02 & 0.15 \\
NGC2366 & -16.8 & 7.77 & 1.29  & 14.7 & 4.27 &  43.89 & 1.21 \\
NGC3738 & -17.1 & 69.88 & 0.6 & 30.8 & 1.05 & 2132.36 & 0.45 \\
\hline
\end{tabular}
\label{table:Properties}
\end{center}
\tablefoot{\\
\tablefoottext{a}{$\mu_0$ and $M$ represent the central surface brightness and absolute magnitude, respectively, in a certain color band.}\\
\tablefoottext{b}{\citealt{mancera2020robust,shi2021cuspy,kong2022odd}}\\
\tablefoottext{c}{\citealt{de1996h,2001deBlokMassmodelLSB}}\\
\tablefoottext{d}{\cite{oh2015high}}}
\end{table*}

\section{Output parameters for the galaxy modelling}
The modelled-obtained stellar scale height h$_z$ and the stellar radial-to-vertical velocity dispersion $\sigma_r/\sigma_z$ at 1.5R$_D$ are listed in Table \ref{table:output}.
\begin{table*}[h!]
\begin{center}
\hfill{}
\hspace{-0.2cm}
\caption{The model-obtained stellar scale heights $h_{z,*}$ and the values of $\sigma_r/\sigma_z$ at 1.5R$_D$ for the UDGs, LSBs and dIrrs.}

\begin{tabular}{l l l l l | l l l l l l | l l l l l}
\hline
Galaxy && $h_{z,*}$ (kpc) & $\sigma_r/\sigma_z$ &&& Galaxy && $h_{z,*}$ (kpc) & $\sigma_r/\sigma_z$ &&& Galaxy && $h_{z,*}$ (kpc) & $\sigma_r/\sigma_z$ \\
\hline
\hline
UDG &&  & &&& LSB && & &&& dIrr && & \\
\hline
AGC114905 && 1.27 & 1.95 &&& F563-1 && 0.42 & 3.01 &&& CVnIdwA && 0.08 & 2.68 \\
AGC122966 && 0.59 & 2.76 &&& F563-V2 && 0.17 & 2.99 &&& DDO52 && 0.17 & 2.74 \\
AGC219533 && 1.10 & 2.34 &&& F568-3 && 0.25 & 2.89 &&& DDO70 && 0.08 & 1.68 \\
AGC242019 && 1.18 & 2.88 &&& F568-V1 && 0.34 & 3.02 &&& DDO101 && 0.17 & 2.96 \\
AGC248945 && 1.94 & 1.78 &&& F574-1 && 0.42 & 2.5 &&& DDO216 && 0.17 & 2.39  \\
AGC334315 && 0.51 & 2.68 &&& F579-V1 && 0.25 & 2.96 &&& NGC2366 && 0.34 & 2.8 \\
AGC749290 && 1.61 & 2.04 &&& F583-4 && 0.17 & 2.74 &&& NGC3738 && 0.47 & 2.74 \\
\hline
\end{tabular}
\label{table:output}
\end{center}
\end{table*}
\end{appendix}
\end{document}